\documentstyle[12pt,aps,preprint]{revtex}
\tightenlines
\begin{document}

\title{Magnon delocalization in ferromagnetic chains with long-range correlated disorder}

\author{Rodrigo P.A.  Lima and Marcelo L.Lyra}
\address{Departamento de F\'{\i}sica,
Universidade Federal de Alagoas,\\ 57072-970 Macei\'o - AL, Brazil}
\author{ Elton M. Nascimento and Ant\^onio D. de Jesus}
\address{Departamento de F\'{\i}sica, Universidade Estadual de Feira de
Santana,\\
44031-460 Feira de Santana - BA, Brazil}

\maketitle

\begin{abstract}

We study one-magnon excitations
in a random ferromagnetic Heisenberg chain with
long-range correlations in the coupling constant distribution.
By employing an exact diagonalization procedure, we compute the localization
length
of all one-magnon states within the band of allowed energies $E$.
The random distribution of coupling constants was assumed to have a
power spectrum decaying as $S(k)\propto 1/k^{\alpha}$. We found that for $\alpha <
1$, one-magnon excitations remain exponentially localized with the localization
length $\xi$ diverging as $1/E$. For $\alpha = 1$ a faster
divergence of $\xi$ is obtained. For any $\alpha > 1$, a phase
of delocalized magnons emerges at the bottom of the band. We characterize the
scaling behavior of the localization length on all regimes and relate it
with the scaling properties of the long-range correlated exchange coupling
distribution.\\
PACS: 75.10.Jm, 75.30.Ds, 75.50.Lk
\end{abstract}
\newpage

\section{introduction}

The properties of quasi-particle excitations in low-dimensional
disordered systems has been the subject of recent intensive investigations due
to the possible applications of random structures and super-lattices in new devices.
In many particular cases of interest, one-dimensional Hamiltonians can incorporate the
main aspects of the disorder in what concerns  the spatial distribution
of quasi-particle excitations, such as optical, acoustic, electronic and
 spin-waves.
Scaling theory establishes that in one-dimensional systems with uncorrelated disorder,
the characteristic length of excitations is finite for any amount of disorder as a consequence
of exponentially localized states\cite{abrahams}.  Correlations in the disorder distribution can stabilize
extended states. Resonant extended states emerge, for example, in one-dimensional electronic
systems with randomly distributed impurity segments\cite{flores1,dunlap,wu} and
interpenatrating Anderson chains\cite{lazo}. Recently, it has been demonstrated that
long-range correlations in the disorder distribution can be responsible for the
emergence of a phase of extended states within the band of allowed energies,
with mobility edges  separating localized and extended states\cite{chico,izrailev}. A recent optical experiment
has demonstrated this phenomenon\cite{apl} which has been proposed for use in the
development of window filters in
electronic, acoustic, or photonic non-periodic structures\cite{izrailev2}.

Spin-wave excitations in ferromagnetic chains with  randomly distributed exchange
couplings $J$ have similar features as those of
electronic excitations in chains with a particular distribution of pair-correlated off-diagonal
 disorder\cite{evangelou1,ziman,theodorou,riklund,evangelou2}.
The localization length of the low-energy one-magnon excitations diverges as
$\xi (E) \propto E^{-\phi}$, where $\phi$ depends
on the particular form of the disorder distribution in the vicinity of $J=0$\cite{ziman,theodorou}.
A diverging length of low-energy excitations is also observed in disordered harmonic chains\cite{harmonic1,harmonic2}.
In both cases, this behavior is associated with the presence of an
uniformly ordered ground state.

Long-range correlations in electronic systems with off-diagonal disorder
can stabilize extended states in a more effective way than in systems
with diagonal disorder\cite{chico2}. As the magnon equations of motion in ferromagnetic
spin chains can be mapped onto those of electronic chains with off-diagonal disorder, one would expect
 the one-magnon excitations in disordered ferromagnetic chains
to be also quite sensitive to the presence of long-range correlations in the
random exchange distribution. In this paper we study, via direct diagonalization and
finite size scaling, the nature of the one-magnon eigenstates in a disordered
chain with long-range correlated disorder with power spectrum decaying as
$S(k)\propto k^{-\alpha}$. Our results indicate that a phase of
extended magnons emerges at the bottom of the band for $\alpha >1$. We characterize
the scaling aspects of the localization length and relate it with those of the disorder
distribution.

\section{Model, formalism and numerical procedure}

The Hamiltonian describing a spin $1/2$ quantum Heisenberg ferromagnetic chain with random exchange
couplings has the form
\begin{equation}
H = -\sum_{n=1}^L J_n\hat{\bf S}_n\cdot\hat{\bf S}_{n+1}~~~,
\end{equation}
the ground state of the chain being the ordered state with all spins aligned.
The one-magnon excitations have the general form $|\Psi\rangle = \sum_n b_n|n\rangle$, where $|n\rangle$ represents
a state with the spin on the $n$-th site flipped with respect to the ground state orientation. The
coefficients $b_n$ can be shown to satisfy the difference equation
\begin{equation}
(J_{n-1}+J_n)b_n - J_{n-1}b_{n-1} - J_nb_{n+1} = Eb_n
\end{equation}
where $E$ is the excitation energy. For an uncorrelated
distribution of the exchange couplings, the one-magnon excitations
are exponentially localized in the thermodynamic limite for any
$E>0$. The uniform $E=0$ mode is quite particular since this mode is not sensitive to
the presence of  disorder. This feature is responsible for the emergence of
effectively extended low-energy excitations in finite chains\cite{ziman,theodorou}.
The typical localization length $\xi$ diverges as
$E\rightarrow 0$ with a power-law whose exponent depends on the
specific form of the distribution $P(J)$ at the vicinity of $J=0$.
For distributions with a finite average $\langle 1/J\rangle $, the inverse
localization length vanishes linearly with $E$. However, for
distributions with a diverging $\langle 1/J\rangle $, the excitations with low
energy become more localized and only a slower
divergence  $\xi\propto E^{-\phi}$ takes place with an exponent $\phi < 1$
which depends of the specific behavior of $P(J)$ at small
values of $J$. This transition at $E=0$, which exists even for
uncorrelated disorder, is due to the fact that the ground state is not
affected by the disorder. In contrast, this feature is not present in the usual
electronic problem with uncorrelated disorder.

In this work we will investigate the role played by long-range correlations in
the exchange coupling distributions. In the analog electronic system, it has been recently
demonstrated that long-range correlations can drastically modify the localized character of
the excitations and even induce the emergence of a phase of delocalized states near the center
of the band\cite{chico2}.

In order to introduce long-range correlations in the disorder
distribution, the exchange couplings $J_n$ will be considered
to be in such a sequency to describe the trace of a fractional
Brownian motion with a specified spectral density
$S(k)\propto 1/k^{\alpha}$, where $k$ is related to the
 wavelength $\lambda$ of the ondulations
on the random exchange landscape by $k=1/\lambda$. For $\alpha = 0$, one recovers the
 traditional disordered ferromagnetic chain model with $\delta$-correlated disorder
$\langle J_nJ_{n'}\rangle =
 \langle J^2\rangle\delta_{n,n'}$.  The exponent $\alpha$
  describes the self-similar character of
the random distribution and the persistent character of its increments. Following
 an approach based on the use of
discrete Fourier transforms\cite{feder,osborne,greis}, a power-law spectral density
is imposed by construction whenever the exchange couplings
 are  given by
\begin{equation}
J_n = \sum_{k=1}^{L/2}
 \left[k^{-\alpha }\left(\frac{2\pi }{L}\right)^{(1-\alpha )}\right]^{1/2}
\cos{\left( \frac{2\pi nk}{L} +\phi_k\right)}
\end{equation}
where $L$ is the number of sites and $\phi_k$ are $L/2$ random phases
uniformly distributed in the interval $[0,2\pi ]$. The equation
above is a general decomposition of a power-law correlated
potential where randomness is present only in the relative phases
of the Fourier components.
We normalize the width of the disorder distribution to keep it
finite and size independent with $\langle J^2\rangle -\langle J\rangle^2 =1$.
Such procedure implies that the
couplings generated by Eq.~3 have to be rescaled by $L^{-\alpha/4}$. In what follows, we
 shift the couplings to have $\langle J_n\rangle=4$
 to enforce all generated couplings to be strictly positive. Those rare
sequences not satisfying this bound where not considered on the subsequent
analysis. The typical dependence of the coupling constant
landscape with the spectral density exponent $\alpha$ are similar
to those reported in Ref.\cite{chico}. The most relevant aspect is
that the landscape becomes progressively less rough as $\alpha$ is
increased, favoring therefore to delocalization.

Our numerical procedure consisted in obtain all eigen-energies and one-magnon eigen-states
for finite chains of size $L$ by direct diagonalization and to infer about
the thermodynamic behavior through a finite size scaling analysis. We performed
a configurational average over distinct disorder realizations such that, for each
chain size, we computed $16\times10^4$ eigenstates. From each normalized eigen-state ($\sum_n b_n^2=1$), we computed
the inverse participation ratio $Y = \sum_n b_n^4$ and used it to characterize the localized/delocalized
nature of these one-magnon states. The typical localization length will be estimated by
$\xi = 1/Y$. In  general, the localization length is finite for
exponentially localized states and diverges linearly with $L$ for trully extended states. In the next
section, we report our results for its scaling behavior for distinct long-range correlated coupling
constants characterized by the power spectrum exponent $\alpha$.

\section{results and discussions}

The behavior of the inverse participation ratio $Y$ and the typical localization length $\xi$ as a
function of the one-magnon energy are shown in figure 1a and 1b respectively
as obtained from chains with $L=1600$ sites and several power spectrum exponents $\alpha$.
For $\alpha = 0$, which mimics an uncorrelated disorder distribution, we found that $Y\propto E$
at low-energy excitations, as expected. This trend remains in the range of $\alpha <1$. In this
regime, we also found the localization length to be size independent over the entire band
of allowed energy values (except for finite size scaling corrections near $E=0$), even tough the local
disorder scales down as $L^{-\alpha /4}$.
Our results indicate that the localization length of the excitations are not sensitive to
such rescaling in this regime. In Fig.~2, we plot the normalized
localization length $\xi(E)/L$ for $\alpha = 0.5$ and several
chain sizes to illustrate the comments above. $\xi(E=0)/L$ remains
finite in the thermodynamic limite once the uniform mode remains
delocalized. For any finite energy, the one-magnon excitations are
exponentially localized with $\xi(E)/L\rightarrow 0$ as $L$
increases.

For $\alpha = 1 $, a new scaling behavior was identified where at low energies
 $Y\propto E^{\phi}$, with $\phi > 1$, indicating that
in  this case the excitations have a stronger tendency to become delocalized.
It is important to mention that a similar feature ($\phi > 1$) was reported to be
present in the analog electronic chain model with long-range correlated hopping
 amplitudes\cite{chico2}.
For any $\alpha > 1$, the data suggest that a phase of delocalized low-energy magnon states
emerges,
which is seen quite clear in figure 1b where $\xi /L$ achieves a plateau. Therefore, long-range
correlations can stabilize extended one-magnon states whenever the power spectrum exponent
$\alpha > 1$.

 To further characterize the scaling behavior of the localization length in the regime of strong
 correlations
that allows for the existence of delocalized states, we employed a finite size scaling study of data from chains with
$L=1600,800,400,200$ and $100$ sites. The behavior of $\xi(E)/L$ is shown in figure 3a for
the typical case of $\alpha = 2$.
For $E<E_c=3.0(2)$ the data collapse characterizes a phase with delocalized states. $E_c$ depends
on the correlation exponent $\alpha$ and on the disorder width. The one-magnon eigen-states remain
exponentially localized for $E>E_c$. However, the localization length in this regime
becomes sensitive to the size-dependent rescaling of the coupling constants. We found that
 the localized states for $\alpha = 2$ have $\xi\propto L^{\gamma}$ with $\gamma = 0.42$ as illustrated by the
 collapse of data above $E_c$ in figure 3b.
This size dependence is weaker than the
expected for a similar rescaling of the local couplings in chains
with uncorrelated disorder. The additional factor  is
due to the correlated nature of the disordered potential and its consequently
enforced finite width. The new scaling exponent $\gamma$ depends on the
correlation exponent $\alpha$ in a non-trivial way. In figure 4, we report our estimated
values for $\gamma$. Our results suggest a non-linear relation $\gamma (\alpha)\propto (\alpha-1)^{0.75}$, although
we could not envisage a simple scaling argument to support this fact. In Figure~5, we show typical wave-functions
representing the localized/delocalized nature of one-magnon
excitations within each phase.

\section{summary and conclusions}

In this work, we investigated the nature of one-magnon excitations in ferromagnetic
quantum chains with long-range correlated disorder. We found that a phase of delocalized
magnons emerges whenever the power spectrum exponent characterizing the strength of
 correlations is $\alpha > 1$, in a close relationship with the behavior of the analog electronic
model with off-diagonal disorder. Chains with
distinct sizes and long-range correlations in the coupling
exchange distribution need to have the local disorder  scaled
down as the size $L$ increases in order to keep the disorder and band width finite.
For $\alpha < 1$, the localized spin excitations are
insensitive to such local disorder rescaling. On the other side, for $\alpha > 1$, the
localized states acquire a size dependent localization length reflecting its sensitivity to
the local disorder rescaling. However, such states should not be confused with critical
multifractal states whose participation ratio also have a power-law size dependence\cite{6a,6b,6c,6d}. The
present localized states have exponential tails and their anomalous scaling is linked to
the scaling property of the underlying long-range correlated potential. It would be
valuable to investigate the spin-wave dynamics in these new regimes
and the possible
relation between super-diffusion and correlation exponents. This information would be
valuable for possible future spin-wave dynamics experiments on correlated
ferromagnetic chains and/or non-periodic ferromagnetic super-lattices.

\section{acknowledgements}

The authors acknowledge the partial financial support of CNPq and CAPES (Brazilian
research agencies) and FAPEAL (Alagoas State agency). MLL is grateful to F.A.B.F. de Moura
and M.D. Coutinho-Filho for stimulating discussions.

\newpage

\section{Figure Captions}

\noindent
{\bf Figure 1} - (a) The Inverse participation ratio as a function of the one-magnon
energy $E$ for a chain with $L=1600$ spins. From top to bottom $\alpha = 0.0, 1.0, 1.5$.
The linear dependence $Y\propto E$ for low-energy excitations observed for $\alpha = 0$
is typical of uncorrelated potentials with finite $<1/J>$. For $\alpha > 1$ a phase
of low-energy delocalized excitations emerges.\\
(b) The scaled localization length $\xi /L$ versus $E$ for chains with $L=1600$.
From bottom to top $\alpha = 0.0, 1.0, 1.5, 2.0$. The phase of delocalized states appears
as a plateau for $\alpha > 1$.

\noindent
{\bf Figure 2} The scaled localization length $\xi /L$ versus $E$ for $\alpha = 0.5$. From top to
bottom $L=100, 200, 400, 800, 1600$. For excitations with finite energies,
$\xi(L)/L\rightarrow 0$ as $L$ increases, characterizing localized states. The
uniform mode at $E=0$ remains delocalized.

\noindent
{\bf Figure 3} - (a) The scaled localization length $\xi /L$ versus $E$ for $\alpha = 2.0$. From top to
bottom $L=100, 200, 400, 800, 1600$. The phase of delocalized states appears as a
size independent plateau. The critical energy separating localized and delocalized states
is $E_c = 3.0 \pm 0.2$.\\
(b) The scaled localization length $\xi / L^{0.42}$ versus $E$ for $\alpha = 2.0$ showing a data
collapse for $E>E_c$.
From bottom to top $L = 100, 200, 400, 800, 1600$.
The size dependence of the characteristic length of such exponentially
 localized states reflects their sensitivity  to the rescaling of the local disorder.

\noindent
{\bf Figure 4} - Estimated values of the exponent $\gamma$ related to the anomalous scaling of the
localization lenght above $E_c$ for several values of the
correlation exponent $\alpha$. The data suggests
$\gamma(\alpha)\propto(\alpha-1)^{0.75}$.

\noindent
{\bf Figure 5} - Typical wave-functions representing delocalized
and localized one-magnon states of a long-range correlated
ferromagnetic chain with $\alpha = 1.75$. (a) Delocalized state
near $E=2.0$; (b) Localized state near $E=8.0$.

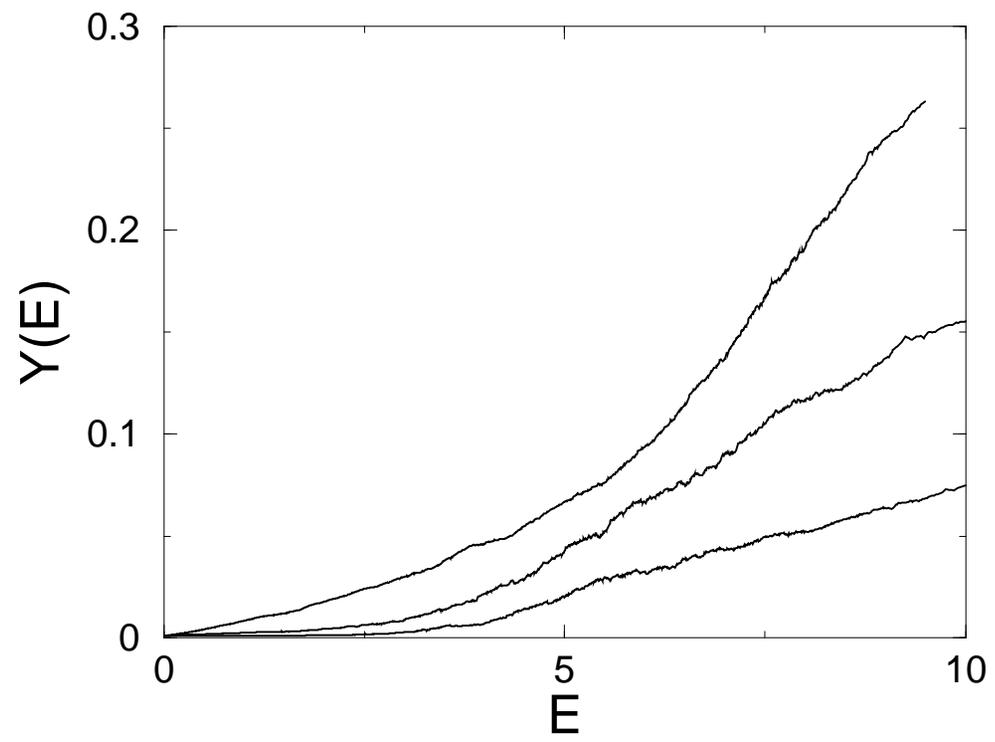
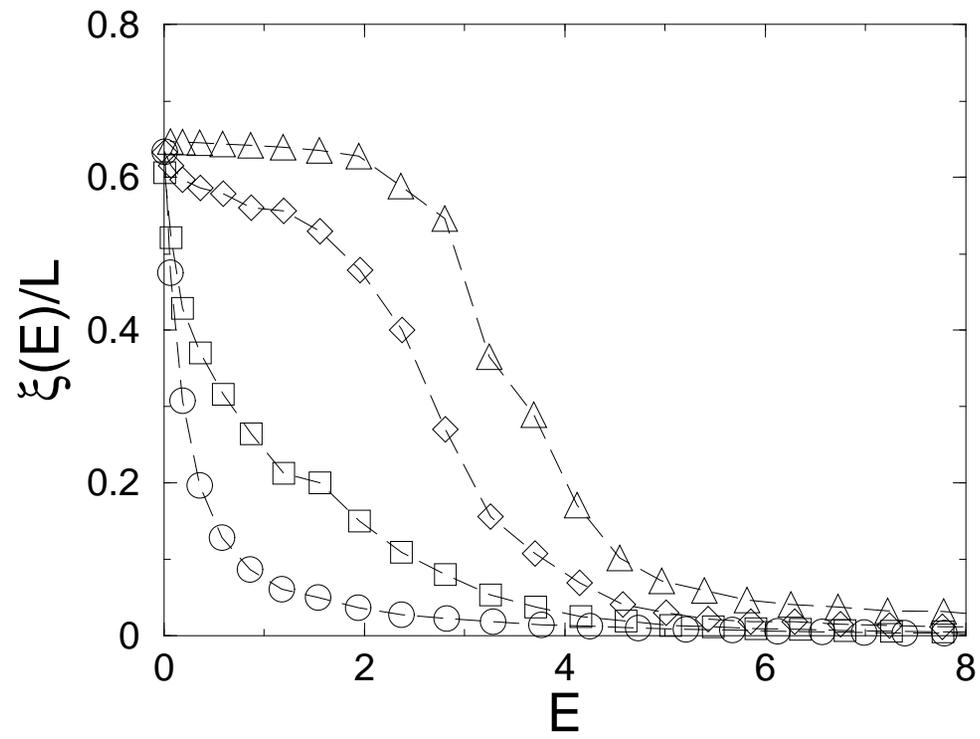
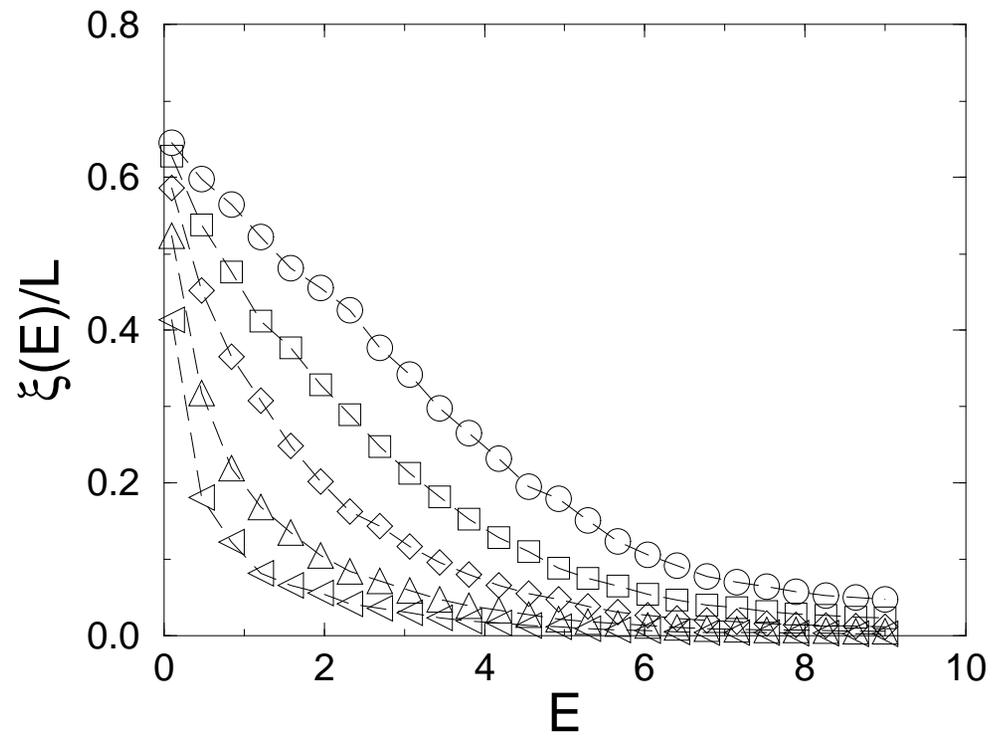
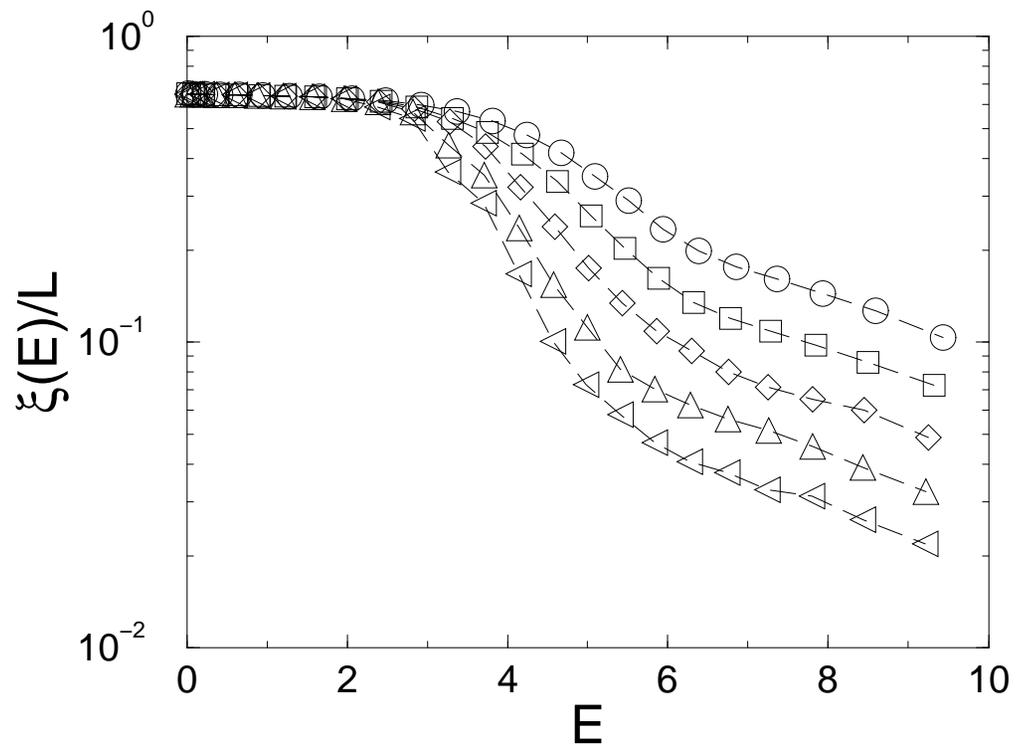
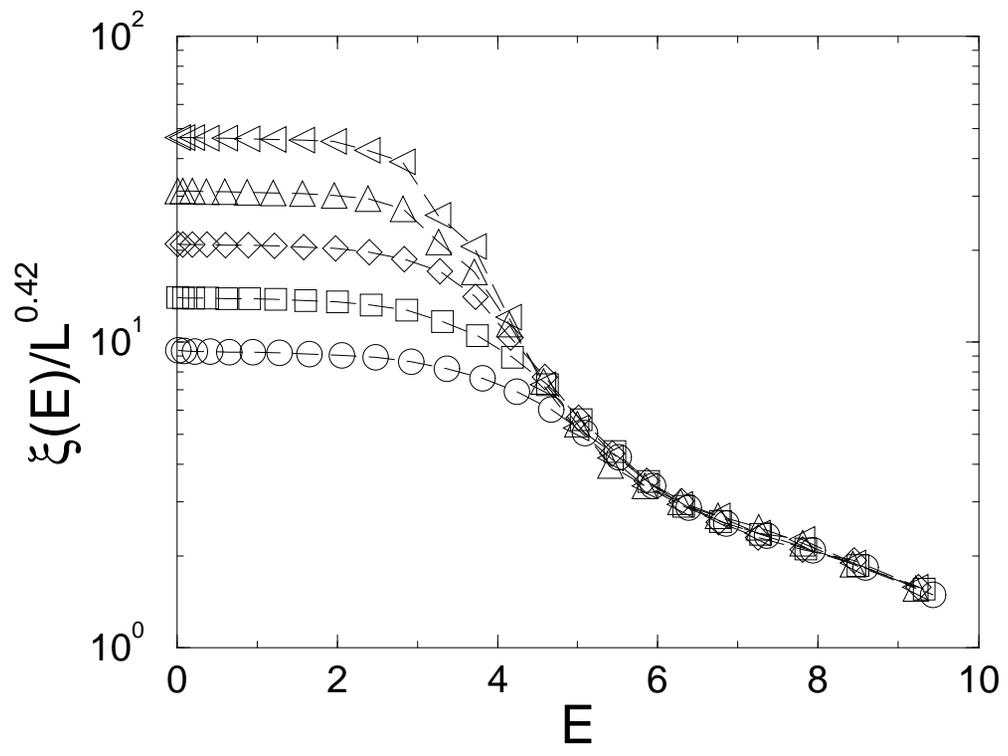
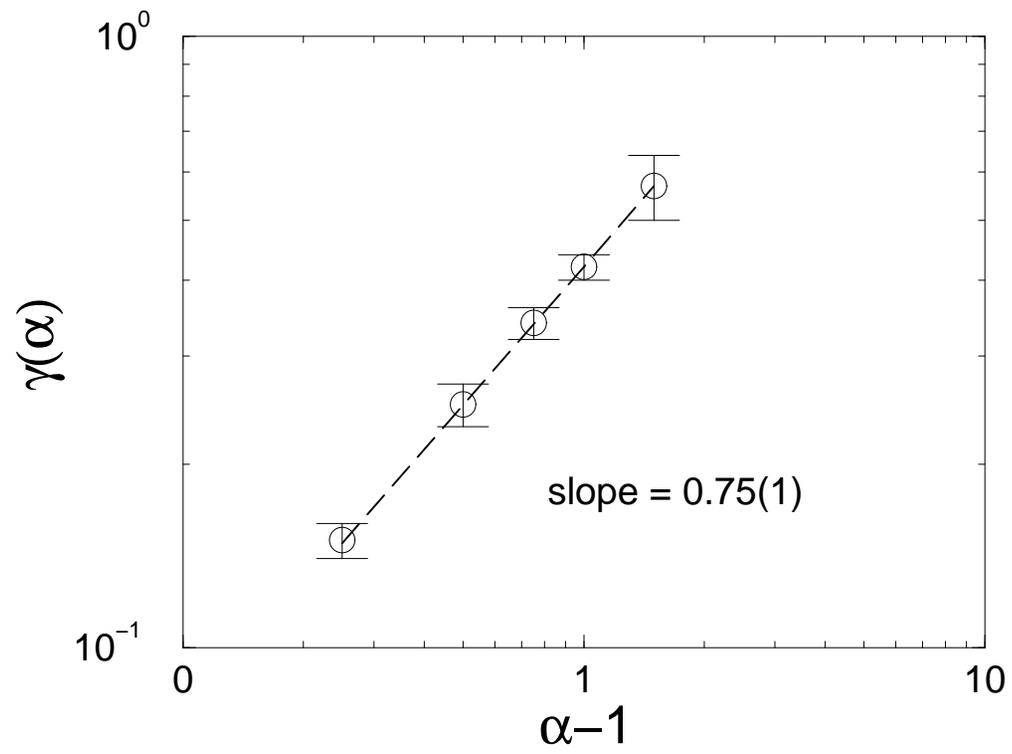
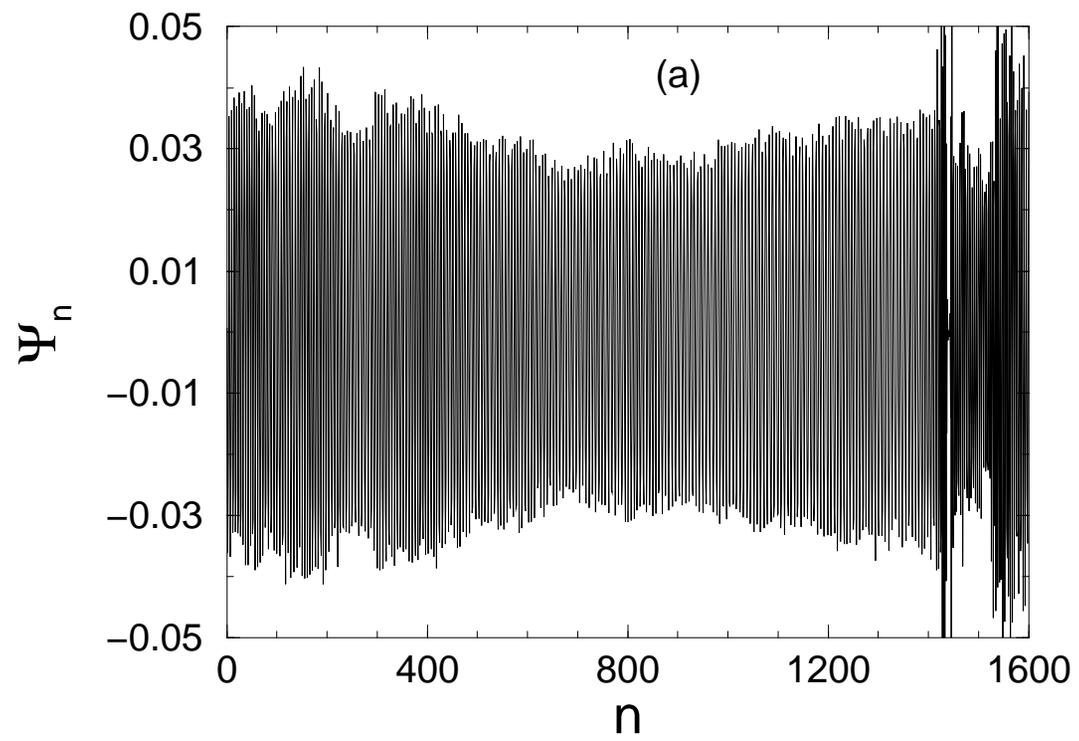
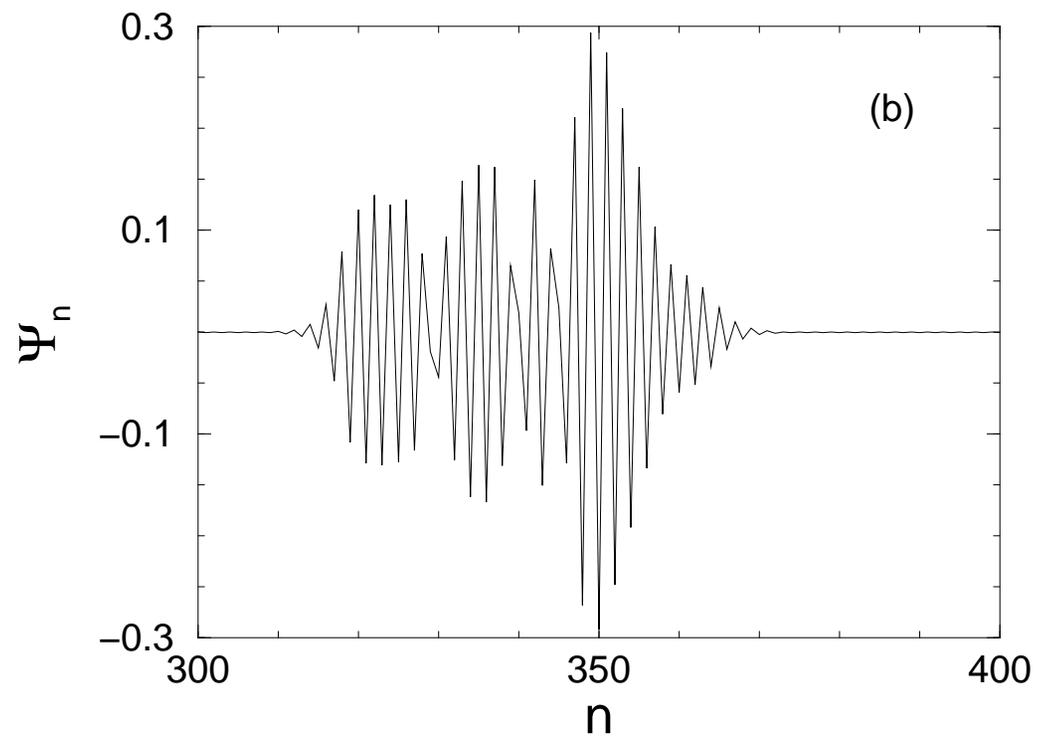
\end{document}